\documentclass[aps,prl,showpacs,twocolumn,superscriptaddress]{revtex4}
\usepackage[utf8x]{inputenc}
\usepackage{graphicx,subfigure,color}
\usepackage{latexsym,amsmath,amssymb,amsfonts,amsthm,enumerate}
\usepackage{amscd}
\usepackage{bm}
\usepackage{wasysym}
\usepackage{pict2e}
\usepackage{float}

\begin{document}
\bibliographystyle{apsrev}

\title{Entropy production in Gaussian bosonic transformations\\ using the replica method: application to quantum optics}

\author{C. N. Gagatsos}
\affiliation{Quantum Information and Communication, Ecole polytechnique de Bruxelles, Universit\'e libre de Bruxelles, 1050 Brussels, Belgium}
\author{A. I. Karanikas}
\affiliation{Nuclear and Particle Physics Section, Physics Department, University of Athens, Panepistimiopolis Ilissia, 15771 Athens, Greece}
\author{G. Kordas}
\affiliation{Nuclear and Particle Physics Section, Physics Department, University of Athens, Panepistimiopolis Ilissia, 15771 Athens, Greece}
\author{N. J. Cerf}
\affiliation{Quantum Information and Communication, Ecole polytechnique de Bruxelles, Universit\'e libre de Bruxelles, 1050 Brussels, Belgium}

\date{\today}

\begin{abstract}
In spite of their simple description in terms of rotations or symplectic transformations  in phase space, quadratic Hamiltonians such as those modeling the most common Gaussian operations on bosonic modes remain poorly understood in terms of entropy production. For instance, determining the von Neumann entropy produced by a Bogoliubov transformation is notably a hard problem, with generally no known analytical solution. Here, we overcome this difficulty by using the replica method, a tool borrowed from statistical physics and quantum field theory. We exhibit a first application of this method to the field of quantum optics, where it enables accessing entropies in a  two-mode squeezer or optical parametric amplifier. As an illustration, we determine the entropy generated by amplifying a binary superposition of the vacuum and an arbitrary Fock state, which yields a surprisingly simple, yet unknown analytical expression.
\end{abstract}

\pacs{03.65.-w, 42.50.-p, 03.67.-a, 89.70.cf}
\maketitle

Gaussian transformations are ubiquitous in quantum physics, playing a major role for instance in quantum optics, quantum field theory, solid state physics, or black hole physics \cite{leonhardt}. In particular, the Bogoliubov transformations resulting from quadratic (bilinear) Hamiltonians in bosonic mode operators are among the most significant Gaussian bosonic transformations, well known to model superconductivity \cite{Bogoliubov} but also describing a much wider range of physical situations, from squeezing or parametric down-conversion in the context of quantum optics \cite{Loudon80,Loudon87,Slusher85,Wu86} to Unruh radiation in an accelerating frame \cite{ful73,davies75,unruh76} or even Hawking radiation as emitted by a black hole \cite{hawk74,hawk75,adami14}.

In order to be concrete,  we focus on Gaussian transformations in quantum optics, which are at the heart of so-called Gaussian quantum information theory \cite{wee12}. For instance, the coupling between two modes of the electromagnetic field as effected by a beam splitter in bulk optics or an optical coupler in fiber optics is modeled by the (passive) quadratic Hamiltonian $H = i \hat a \hat b^\dagger  - i \hat a^\dagger \hat b$, where $\hat a$ and $\hat b$ are bosonic mode operators. This operation can be shown to preserve the Gaussian character of a quantum state, or more precisely the quadratic exponential form of its characteristic function. The corresponding transformation in phase space is the rotation 
$\hat a \to \cos \theta \, \hat a + \sin \theta \, \hat b$ and $\hat b \to \cos \theta \, \hat b - \sin \theta \, \hat a$,  where $\cos^2\theta$ is the transmittance.
Another generic Gaussian coupling between two modes of the electromagnetic field 
results from parametric down-conversion in a nonlinear medium, which is modeled by the (active)
quadratic Hamiltonian $H =  i \hat a \hat b  -i  \hat a^\dagger \hat b^\dagger$. It effects the Bogoliubov transformation 
$\hat a \to \cosh r \, \hat a + \sinh r \, \hat b^\dagger$ and $\hat b \to \cosh r \, \hat b + \sinh r \, \hat a^\dagger$, where $\cosh^2 r$ is the parametric amplification gain, and is traditionally used as a source of quantum entanglement. More generally, the set of linear canonical transformations effected by quadratic bosonic Hamiltonians, also referred to as symplectic transformations, can easily be characterized in terms of affine transformations in phase space (e.g., rotations, area-preserving squeeze mapping).

Unfortunately, while they are common, these quantum Gaussian processes are poorly understood in terms of entropy generation. Indeed, the phase-space representation is not suited to calculate von Neumann entropies, which requires diagonalizing density operators in state space \cite{Wehrl}. For example, when  amplifying an optical state using parametric down-conversion, the output state suffers from quantum noise, which is an increasing function of the amplification gain \cite{Caves}. It is a central problem to characterize this noise via the entropy of the output state, this being indispensable in particular to determine the capacity of Gaussian bosonic channels \cite{hol01,giov04,Raul12a}. The output entropy is, however, not accessible for an arbitrary input state because it is difficult -- usually impossible -- to  diagonalize the corresponding output state in an infinite-dimensional Fock space. With the exception of Gaussian states, e.g., the vacuum state (resulting after amplification in a thermal state whose entropy is given by a well-known formula), very few analytical results are available as of today \cite{Raul12a}.

In this Letter, we demonstrate that the \emph{replica method} can be successfully exploited in order to overcome this problem and find the exact analytical  expression of the output entropy of Gaussian processes acting on some non-trivial bosonic input states. The replica method is well known to be a very useful tool in statistical physics, especially with disordered systems \cite{replica-book}, and in quantum field theory \cite{callan94}. Here, we first apply it to the field of quantum optics and show that it enables accessing the entropy generated by a quantum optical amplifier, opening a new way towards the entropic characterization of Gaussian transformations generated by quadratic bosonic Hamiltonians. To illustrate the power of this approach in a non-trivial case, we calculate the output entropy when amplifying a binary superposition of the vacuum and an arbitrary  Fock state, which yields a surprisingly simple analytical expression.

\emph{Replica method.}---Calculating the von Neumann entropy $S(\hat\rho)=-\textrm{tr}(\hat{\rho} \ln \hat{\rho})$ of a bosonic mode that is found in state $\hat \rho$ is often an intractable task because it requires finding the infinite vector of eigenvalues of $\hat\rho$. This can sometimes be circumvented by using the replica method, which relies on the identity $\log Z = \lim_{n\to 0^\mathrm{+}} (Z^n-1)/n$. 
Using $x \log x = \lim_{n \to 1^\mathrm{+}} (x^n-x) / (n-1) = \frac {\partial} {\partial n} (x^n)\big|_{n=1^\mathrm{+}}$, we may reexpress the von Neumann entropy as
\begin{equation}  \label{eq-replicarep}
S(\hat \rho)=-\frac{\partial \, \textrm{tr}(\hat \rho^n)}{\partial n}\Big|_{n=1^\mathrm{+}}
\end{equation}
The trick is to find an analytical expression of $\textrm{tr}(\hat \rho^n)$ as a function of $n \in \mathbb{N}^*$ and to compute its derivative at $n=1$, avoiding the need to diagonalize $\hat \rho$. This method also makes apparent the connection between the von Neumann entropy and other widely used measures of disorder,
such as Tsallis and R\'enyi entropies. It has been used with great success in the context of spin glasses 
and quantum field theory \cite{replica-book,callan94,hol94,calab04-05,ryu06,berger08,gag13}, being justified based on the analyticity of $\textrm{tr}(\hat{\rho}^n)$ in a neighborhood of $n$=$1$ \cite{calab04-05}. In the Supplemental Material \cite{SM}, we connect it with Hausdorff's moment problem and provide some easy but instructive examples from classical probability theory.

As we shall show in this Letter, dealing with quadratic interactions makes the replica method an invaluable tool to access the von Neumann entropy because it involves Gaussian integrations, or else tricks can be used in order to bring $\textrm{tr}(\hat \rho^n)$ to a calculable form. Let us illustrate this principle with a generic zero-mean rotation-invariant Gaussian state, namely a thermal state $\hat \rho_0 = (1-|\tau|^2) \textstyle\sum_{k=0}^\infty |\tau|^{2k} \, |k\rangle\langle k|$ characterised by a mean photon number $N=|\tau|^2/(1-|\tau|^2)$.
Since $\hat \rho_0$ is in a diagonal form, it is of course straightforward to calculate its entropy, giving the well-known formula $S(\hat \rho_0)=g(N)\equiv (N+1)\log(N+1)-N\log N$. However, we may also start with its non-diagonal representation in the coherent-state basis $\{|\alpha\rangle\}$, where $\alpha$ is a complex number, namely
\begin{eqnarray}
\hat \rho_0=\frac{1}{\pi}\frac{1-|\tau|^2}{|\tau|^2}
\int d^2 \!\alpha \, e^{-\frac{1-|\tau|^2}{|\tau|^2}|\alpha|^2}|\alpha\rangle\langle \alpha|
\end{eqnarray}
By making the change of variable $\alpha \to |\tau| \ \alpha$ and by using $\langle \alpha | \beta \rangle = e^{-(|\alpha|^2+|\beta|^2-2 \alpha^*\beta)/2}$, we can write
\begin{eqnarray}
\textrm{tr}(\hat\rho_0^n)=\frac{(1-|\tau|^2)^n}{\pi^n}
\int d^2 \!\alpha_1 \ldots \int d^2 \!\alpha_n \, e^{-\bar\alpha^\dag M \bar\alpha}
\label{thermal}
\end{eqnarray}
where $\bar\alpha=(\alpha_1,\ldots \alpha_n)^T$ is a column vector  and $M$ is the $n\times n$ circulant matrix
\begin{eqnarray}
M=
\begin{pmatrix}
1             & -|\tau|^2 & 0         & \ldots & 0\\
0             & 1         & -|\tau|^2 & \ldots & 0\\
\vdots        & \vdots    & \vdots    & \ddots    & \vdots\\
-|\tau|^2     & 0     & 0         & \ldots & 1
\end{pmatrix}.
\label{M}
\end{eqnarray}
Equation (\ref{thermal}) is a simple Gaussian integral, which, using the determinant $\det M = 1-|\tau|^{2n}$, can be expressed as
\begin{eqnarray}
\textrm{tr}(\hat\rho_0^n)=\frac{(1-|\tau|^2)^n}{1-|\tau|^{2n}}
\label{thermalMom}
\end{eqnarray}
Then, we readily find that
\begin{eqnarray}
-\frac{\partial}{\partial n} \textrm{tr}(\hat\rho_0^n)\Big|_{n=1}=
\ln\frac{1}{1-|\tau|^2}+\frac{|\tau|^2}{1-|\tau|^2}\ln\frac{1}{|\tau|^2}
\label{thermalEntr}
\end{eqnarray}
which coincides with the above expression $S(\hat \rho_0)=g(N)$ for the entropy of a thermal state, as expected.

\emph{Amplifying a Fock state.}---Consider now the harder problem of expressing the entropy $S_m$ generated by amplifying an arbitrary Fock state $|m\rangle$. Thus, we consider a two-mode squeezer of parameter $\xi=|\xi|\, e^{i \phi}$, applying the unitary transformation
\begin{eqnarray}
\hat{U}=e^{-\xi \hat{a}^{\dag} \hat{b}^{\dag}+\xi^* \hat{a} \hat{b}}
\label{tms}
\end{eqnarray}
on the initial state $|m\rangle_a |0\rangle_b$ (subscript $a$ refers to the signal mode, while $b$ refers to the idler mode). 
The reduced output state $\hat{\rho}_m$ of the signal mode is diagonal in the Fock basis, with a vector of eigenvalues given by \cite{Raul12a,Raul12b}
\begin{eqnarray}
p^{(m)}_k&=&(1-|\tau|^2)^{m+1} \binom{k+m}{k} |\tau|^{2k}  , \quad  k \in \mathbb{N}
\label{eigm}
\end{eqnarray}
where $\tau = \tanh \xi$, from which we find
\begin{eqnarray}
\textrm{tr}(\hat\rho^n_m )=(1-|\tau|^2)^{n(m+1)}   \sum_{k=0}^{\infty} \binom{k+m}{k}^n |\tau|^{2nk}.
\label{summ}
\end{eqnarray}
Equation (\ref{summ}) can be re-expressed in a closed form as
\begin{eqnarray}
\textrm{tr}(\hat\rho^n_m )=\frac{(1-|\tau|^2)^{n(m+1)}}{|\tau|^{2n}} \, \textrm{Li}_{-n}^{(m)}(|\tau|^{2n}).
\label{trmn}
\end{eqnarray}
where
\begin{eqnarray}
\textrm{Li}_{-n}^{(m)}(\zeta)  \equiv  \sum_{k=0}^{\infty} \binom{k+m}{k}^n \zeta^{k+1}
\label{Lim}
\end{eqnarray}
and $\textrm{Li}^{(1)}_{-n}(\zeta)$ denotes the polylogarithm of order $-n$ \cite{handbook}.
Applying Eq. (\ref{eq-replicarep}) to Eq. (\ref{trmn}) and taking into account that $\textrm{Li}_{-1}^{(m)}(\zeta)=\zeta/(1-\zeta)^{m+1}$, we obtain the entropy
\begin{eqnarray}
\lefteqn{   S_m=-\frac{\partial}{\partial n} \textrm{tr}(\hat\rho^n_m)\Big|_{n=1}=\ln \frac{|\tau|^2 ~~~}{(1-|\tau|^2)^{m+1}}     }\nonumber \\
& ~~~~~~~~~~~   -\frac{(1-|\tau|^2)^{m+1}}{|\tau|^2 ~~~}\frac{\partial}{\partial n} \, \textrm{Li}_{-n}^{(m)}(|\tau|^{2n})\Big|_{n=1}.
\label{Sm}
\end{eqnarray}
Using $\textrm{Li}_{-n}^{(0)}(\zeta)=\zeta/(1-\zeta)$, it is easy to check that Eq.~(\ref{Sm}) gives the correct value for $S_0$, that is, the entropy of a thermal state $\hat\rho_0$ as in Eq.~(\ref{thermalEntr}). Also, a closed expression can be found for $S_1$ in terms of Eulerian numbers as the polylogarithm $\textrm{Li}_{-n}^{(1)}(\zeta)$ is a well studied function, see \cite{SM}. For $m>1$, the function $\textrm{Li}_{-n}^{(m)}(\zeta)$ assumes a summation form which is convergent and differentiable with respect to $n$, yielding an analytical expression for $S_m$, see \cite{SM}.

\emph{Superposition of Fock states.}---We will now show that the same procedure makes it possible to express the entropy analytically in situations where no diagonal form is available for the output state, so the replica method becomes essential. Consider the amplification of a binary superposition of the type 
\begin{eqnarray}
|\psi \rangle=\frac{|0\rangle+z |m\rangle}{\sqrt{1+z^2}}
\label{initial}
\end{eqnarray}
where we take $z \in \mathbb{R}$ without loss of generality. 
By using the Baker-Campbell-Hausdorff relation, the unitary transformation (\ref{tms}) can be rewritten in the form
\begin{eqnarray}
\hat{U}=\textrm{e}^{-\nu}\textrm{e}^{-\tau \hat{a}^{\dag} \hat{b}^{\dag}}
\textrm{e}^{-\nu(\hat{a}^{\dag} \hat{a}+\hat{b}^{\dag} \hat{b})}
\textrm{e}^{\tau^* \hat{a} \hat{b}}
\label{tms op}
\end{eqnarray}
where $\nu=\ln \cosh |\xi|$ and $\tau=\frac{\xi}{|\xi|}\tanh|\xi|$, so that the joint output state $|\Psi \rangle$ of the two modes can be expressed in the double coherent-state basis $|\alpha\rangle_a |\beta\rangle_b$, namely
\begin{eqnarray}
\nonumber \langle\alpha, \beta| \Psi \rangle &=&\frac{1}{\sqrt{1+z^2}}\left(\langle\alpha,\beta|\hat{U}|0,0\rangle + z \, \langle \alpha,\beta |\hat{U}|m,0\rangle\right) \\
\nonumber &=& \frac{1}{\sqrt{1+z^2}}\left( (1-|\tau|^2 )^{\frac{1}{2}}+z\frac{(1-|\tau|^2 )^{\frac{m+1}{2}}} {\sqrt{m!}}\right) \\
&&  \times e^{-(|\alpha|^2+|\beta|^2)/2-\tau \alpha^*\beta^*+\tau^* \alpha \beta}. 
\label{final state}
\end{eqnarray} 
From Eq.~(\ref{final state}), we can easily write the reduced output state $\hat\rho$ obtained by tracing $|\Psi \rangle$ over the idler mode and paying attention to the non-orthogonality of coherent states. Using the notation $\bar\alpha=(\alpha_1,\ldots \alpha_n)^T$, we get
\begin{eqnarray}
\textrm{tr} (\hat{\rho}^n)=\frac{(1-|\tau|^2)^n}{\pi^n (1+z^2)^n} \prod_{j=1}^n  \int  d^2 \!\alpha_j  \left| 1+c \alpha_j^m \right|^2 
e^{-\bar\alpha^\dag M \bar\alpha}
\label{trn}
\end{eqnarray}
where the matrix $M$ is defined as in Eq.~(\ref{M}) and
\begin{eqnarray}
c=\frac{z}{\sqrt{m!}}(1-|\tau|^2)^{m/2}.
\label{c}
\end{eqnarray}
In order to bring this back to a Gaussian integral, we use the so-called ``sources'' trick \cite{source}, exploiting the identity $x^m e^{-x^2} = \frac{\partial^m}{\partial\lambda^m} e^{-x^2+\lambda x}|_{\lambda=0}$.
Then, Eq. (\ref{trn}) becomes
\begin{eqnarray}
\nonumber \textrm{tr} (\hat{\rho}^n)&=& \frac{(1-|\tau|^2)^n}{\pi^n (1+z^2)^n}  \, \Pi_{\partial \lambda}(n) \prod_{j=1}^n \int d^2 \!\alpha_j \\
&& \times\exp\left(-\bar\alpha^\dagger M \bar\alpha + \bar\alpha^\dagger \bar\lambda + \bar\lambda^\dagger \bar\alpha \right)\Big|_{\bar\lambda=\bar 0}
\label{trnSources}
\end{eqnarray}
where $ \Pi_{\partial \lambda}(n) \equiv \prod_{j=1}^n \left| 1+c \, \partial^m/\partial\lambda_j^m \right|^2$ is a differential operator 
in the variables  $\bar\lambda= (\lambda_1,\ldots \lambda_n)^T$. Note here that $\lambda_j$ and $\lambda_j^*$ are treated as independent variables, instead of their real and imaginary parts.
The derivatives with respect to all $\lambda$'s have been pushed in front of the integrals in Eq. (\ref{trnSources}), so that we get a Gaussian integral
that is immediately calculable, resulting in
\begin{eqnarray}
\textrm{tr}(\hat{\rho}^n)=\frac{\textrm{tr}(\hat\rho_0^n )}{(1+z^2)^n} \, \Pi_{\partial \lambda}(n)\exp\left(\bar{\lambda}^\dagger N \bar{
\lambda}\right)\Big |_{\bar{\lambda}=\bar{0}} ,
\label{trnPiMain}
\end{eqnarray}
where $\textrm{tr}(\hat\rho_0^n )$  is given by Eq.~(\ref{thermalMom}) and corresponds to a vacuum input state ($z=0$).
In Eq. (\ref{trnPiMain}), we have defined the circulant matrix $N=(1-|\tau|^{2n})^{m} M^{-1}$, with
\begin{eqnarray}
 M^{-1} = \frac{1}{1-|\tau|^{2n}}
\begin{pmatrix}
1                         & |\tau|^2                    & \ldots                    & |\tau|^{2(n-1)}\\
|\tau|^{2(n-1)}           & 1                           & \ldots                    & |\tau|^{2(n-2)}\\
\vdots                    & \vdots                      & \ddots                    & \vdots\\
|\tau|^2                  & |\tau|^4                    & \ldots                    & 1
\end{pmatrix}
\label{C}
\end{eqnarray}
while the operator $\Pi_{\partial \lambda}(n)$ can be expanded as
\begin{eqnarray}
\Pi_{\partial \lambda}(n)=\sum_{k=0}^{n} c^{2k} \, \Pi_{2k}(n)
\label{PiMain}
\end{eqnarray}
where each $\Pi_{2k}(n)$ contains $\binom{n}{k}^2$ terms that return a non-zero result when 
acting on $\exp\left(\bar{\lambda}^\dagger N \bar{\lambda}\right)$ and taking the value at $\bar{\lambda}=\bar{0}$, see \cite{SM} for details. 
The term with $k=0$ in Eq.~(\ref{PiMain}) is simply $\Pi_{0}(n)=1$, so that taking $z=0$ trivially results into $\textrm{tr}(\hat\rho^n )=\textrm{tr}(\hat\rho_0^n )$.
The term with $k=n$ gives, when acting on the exponential of Eq. (\ref{trnPiMain}),
\begin{eqnarray}
\Pi_{2n} \exp\left(\bar{\lambda}^\dagger N \bar{\lambda}\right)\Big |_{\bar{\lambda}=\bar{0}}=
m!^n \frac{1-|\tau|^{2n}}{|\tau|^{2n}} \, \textrm{Li}_{-n}^{(m)}(|\tau|^{2n})
\label{last}
\end{eqnarray}
where $\textrm{Li}_{-n}^{(m)}(|\tau|^{2n})$ is defined in Eq. (\ref{Lim}). Thus, we recognize that this term is connected with the case of an input Fock state $|m\rangle$, something that can 
also be seen by taking the limit $z\rightarrow \infty$ in Eq. (\ref{trnPiMain}).
If we put all pieces together, we obtain the expression
\begin{eqnarray}
\nonumber \lefteqn{    \textrm{tr}(\hat{\rho}^n)=\frac{\textrm{tr}\hat\rho_0^n}{(1+z^2)^n}\Bigg\{ \Bigg[1+  z^{2} \Bigg(\frac{1-|\tau|^2}{1-|\tau|^{2n}}\Bigg)^m \Bigg]^n     }  \\
&& -z^{2n}\Bigg(\frac{1-|\tau|^2}{1-|\tau|^{2n}}\Bigg)^{m n}+F^{(m)}(n)  +z^{2n}\frac{\textrm{tr}\hat\rho_m^n}{\textrm{tr}\hat\rho_0^n}\Bigg\} 
\label{trnFinalMain}
\end{eqnarray}
where $\hat\rho_m$ is the reduced output state resulting from the amplification of $|m\rangle$, and $F^{(m)}(n)$ is defined in \cite{SM}. Now, applying Eq. (\ref{eq-replicarep}) to Eq. (\ref{trnFinalMain}), we get
\begin{eqnarray}
S(z)=\frac{1}{1+z^2} \, S_0+\frac{z^2}{1+z^2} \, S_m-\frac{\partial}{\partial n}F^{(m)}(n)\Big|_{n=1}.
\label{SwithFMain}
\end{eqnarray}
Finally, we prove in \cite{SM} that the last term of the right-hand side of Eq. (\ref{SwithFMain}) vanishes,
so that Eq.~(\ref{SwithFMain}) simplifies into the expression,
\begin{eqnarray}
S(z)=\frac{1}{1+z^2} \, S_0+\frac{z^2}{1+z^2} \, S_m
\label{Sfinal}
\end{eqnarray}
Intriguingly, the output entropy is thus a simple convex combination of $S_0$ and $S_m$ with the exact same weights as if we had lost coherence between the components $|0\rangle$ and $|m\rangle$ of the input superposition. This is schematically pictured in Fig.~1. It is illustrated in Fig.~2, where we show that the entropy is a monotonically increasing function of the superposition parameter $z$.
\\
\begin{figure}[t]
\centering
\includegraphics[scale=0.35]{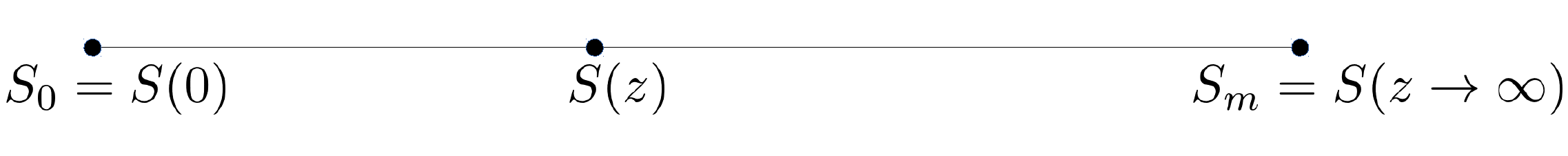}
\caption{The von Neumann entropy $S(z)$,  function of the superposition parameter $z$, is pictured by a point belonging to a one-dimensional convex polytope. The two 
extremal points of the polytope are the entropies $S_0$ and $S_m$ corresponding to the two extreme cases, i.e., the input states $|0\rangle$ and $|m\rangle$.}
\end{figure}

\emph{Conclusions.}---We have demonstrated that the replica method, a tool borrowed from other areas of physics, provides a new angle of attack to access quantum entropies in fundamental Gaussian bosonic transformations (i.e., quadratic interactions between bosonic mode operators such as Bogoliubov transformations). The entropic characteristics of such transformations can be calculated by using the symplectic formalism as long as Gaussian states are considered, but otherwise the problem is generally unsolvable. For instance, it required considerable effort to be able to prove the simplest fact that the vacuum state minimizes the entropy produced by Gaussian bosonic channels \cite{giov14}. The difficulty behind this proof was that no diagonal representation of $\hat{\rho}$ is available, as it is most often the case when non-Gaussian states are considered. This problem is also linked to several unproven entropic conjectures on Gaussian optimality in the context of bosonic channels. Notably, determining the capacity of a multiple-access or broadcast Gaussian bosonic channel is pending on being able to access entropies, see, e.g., \cite{guha08,yen05,guha07}.
The replica method holds the promise to unblock the situation as it provides a trick to overcome this difficulty: $\textrm{tr}(\hat{\rho}^n)$ is expressed for $n$ replicas of state $\hat{\rho}$ by using Gaussian integrals, without ever accessing its eigenvalues.

We have illustrated this procedure with the amplification of a state of the form $|0\rangle+z |m\rangle$. This allowed us to unveil a remarkably simple behavior for the entropy of the amplified state, namely that it is a convex combination of the extremal points $S_0$ and $S_m$. It must be stressed that this analytical result is highly non-trivial as we do not expect similar expressions for the entropy resulting from other superpositions, such as  $|1\rangle+ z |2\rangle$ or $|0\rangle+z |1\rangle+z' |2\rangle$. Take for instance a coherent state: although it is an infinite superposition of Fock states, the resulting entropy is $S_0$, just as for the vacuum state.

In conclusion, we anticipate that the replica method may become an invaluable tool in order to reach a complete entropic characterization of Gaussian bosonic transformations,
or perhaps even solve pending conjectures on Gaussian bosonic channels (in a current work, we further explore this avenue and consider more general Gaussian transformations as well as mixed input states). 
\\
\begin{figure}[t]
\centering
\includegraphics[scale=0.65]{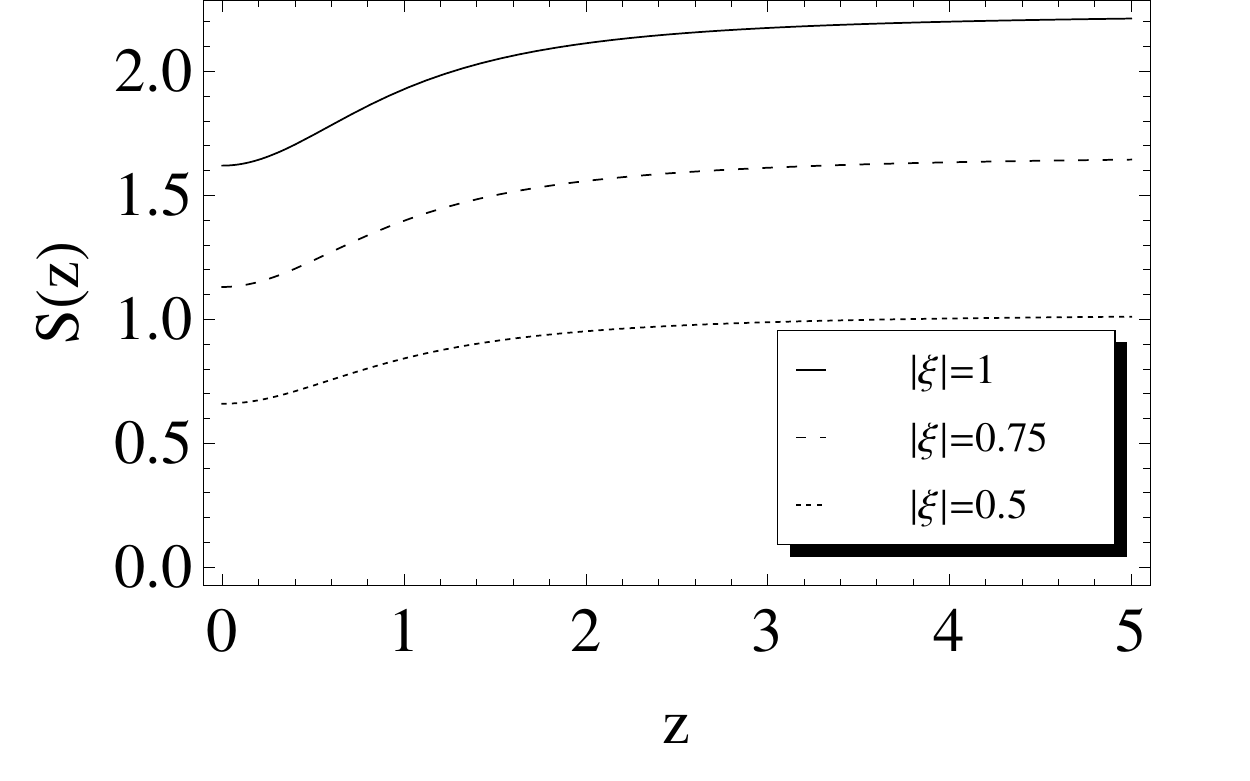}
\caption{Plot of the von Neumann entropy $S(z)$ as a function of the superposition parameter $z$ for $m=1$ and several values of the squeezing parameter  $\xi$. Since $S_1>S_0$, Eq. (\ref{Sfinal}) implies that the curve $S(z)$ is always above $S_0$.}
\end{figure}
\\
\\
\\

\begin{acknowledgments}
We thank R. Garc\'ia-Patr\'on, J. Sch\"afer, and O. Oreshkov for useful discussions. This work was supported by the F.R.S.-FNRS under the ERA-Net project HIPERCOM. C.N.G. 
acknowledges financial support from Wallonia-Brussels International via the excellence grants program.
\end{acknowledgments}

\bibliographystyle{apsrev}

\newpage

~~~

\newpage

\section{Supplemental Material}

\subsection{1. Replica method : justification and examples}

\subsubsection{The replica method}

In quantum mechanics, the von Neumann entropy is defined as
\begin{eqnarray}
S=-\textrm{tr}(\hat{\rho} \ln \hat{\rho})
\label{vne}
\end{eqnarray}
where $\hat{\rho}$ is a density operator. In many cases, this definition is not 
practical as it involves computing the logarithm of a matrix. In other words, we have to find the eigenvalues of the matrix, which is infinite-dimensional for the density operator of a bosonic mode, a task that is  often impossible. One can sometimes circumvent this problem by using the replica method, which is described as follows. We introduce the quantity, 
\begin{eqnarray}
\nonumber \textrm{tr}(\hat{\rho}^n)&=&\int dx_1 \int dx_2 \ldots \int dx_n\\
&&\rho(x_1,x_2)\ldots \rho(x_n,x_1)
\label{repl}
\end{eqnarray}
which can be viewed as the trace of the $n$ replicas of the density matrix $\hat{\rho}$.
Here, we denote by $\rho(x_i,x_j)$ the representation of $\hat{\rho}$ in some continuous basis, although in general the basis that we choose
to represent the density matrix upon does not have to be continuous. In the case of a discrete representation, we would have summations instead of 
integrals in Eq. (\ref{repl}). The Tsallis entropy of order $n$ is a disorder monotone and is defined as,
\begin{eqnarray}
S^{(n)}_T=\frac{1-\textrm{tr}(\hat{\rho}^n)}{n-1}
\label{tsallis}
\end{eqnarray}
It is well known that in the limit $n\to 1$, one recovers the von Neumann entropy, that is,
\begin{eqnarray}
\lim_{n\to 1} S^{(n)}_T=S.
\label{tsallis to vN}
\end{eqnarray}
On the other hand, it holds that,
\begin{eqnarray}
\lim_{n\to 1} \frac{1-\textrm{tr}(\hat{\rho}^n)}{n-1}=-\frac{\partial \textrm{tr}(\hat{\rho}^n)}{\partial n}\Bigg|_{n=1}.
\label{tsallis deriv}
\end{eqnarray}
In numerous papers, arguments of analytic continuation are used to justify that Eq. (\ref{tsallis deriv}) is a correct relation as long as we take the 
derivative at some integer value of $n$, here $n=1$.
Then, from Eqs. (\ref{tsallis to vN}) and (\ref{tsallis deriv}), we find the central equation
\begin{eqnarray}
S=-\frac{\partial \textrm{tr}(\hat{\rho}^n)}{\partial n}\Bigg|_{n=1}.
\label{repmeth}
\end{eqnarray}
The question concerning the justification of the replica method can be rephrased as whether the integer order's moments sequence $\textrm{tr}(\hat{\rho}^n)$ of a density operator $\hat{\rho}$ uniquely determines the density operator $\hat{\rho}$ itself. If this is true, then, using the moments, one could in principle uniquely determine any expectation value, in particular the von Neumann entropy since it is the expectation value $-\mathbb{E}(\ln \hat{\rho})$ in view of Eq. (\ref{vne}). 

\subsubsection{Hausdorff's moment problem}

Let us now recourse to Hausdorff's moment problem in order to justify the replica method. In what follows, we define as moments the quantity $m_n=\textrm{tr}(\hat\rho^n)$. Hausdorff's moment problem states that if $X$ is a random variable in the interval $[0,1]$, then the integer-order moments 
\begin{eqnarray}
m_n \equiv \mathbb{E}(X^{n-1})
\end{eqnarray}
uniquely determine the distribution $P(X)$ if and only if
\begin{eqnarray}
(-1)^k (\Delta^k m)_n\ge 0
\end{eqnarray}
where
\begin{eqnarray}
(\Delta^k m)_n=\sum_{j=0}^{k}(-1)^j \binom{k}{i} m_{n+k-i}.
\end{eqnarray}
In the case at hand, we will consider as random variable  the eigenvalues $\lambda_i$ of the (hermitian) density operator, that is $\Lambda \in [0,1]$ with
\begin{eqnarray}
P(\Lambda=\lambda_i)=\lambda_i
\end{eqnarray} 
The moments,
\begin{eqnarray}
m_n=\mathbb{E}(\hat\rho^{n-1})=\textrm{tr}(\hat\rho^{n-1}\hat\rho)=\textrm{tr}(\hat\rho^n)
\end{eqnarray}
is easily found to satisfy,
\begin{eqnarray}
(-1)^k (\Delta^k m)_n= \sum_j \lambda_j^n(1-p_j)^k > 0.
\end{eqnarray} 
Therefore the knowledge of the moments uniquely determines the density operator in the basis of its eigenvectors, or equivalently the eigenvalues of the density operator. Thus,  we can find in principle any expectation value from the knowledge of $\textrm{tr}(\hat\rho^n)$ for $n \in \mathbb{N}^*$. What we need is a recipe to derive the von Neumann entropy from the moments. An obvious choice comes from observing that if we treat $n$ as a real variable we have,
\begin{eqnarray}
-\frac{\partial \textrm{tr}(\hat{\rho}^n)}{\partial n}\Bigg|_{n=1}=-\textrm{tr}(\hat \rho \ln \hat\rho)=S.
\label{repmeth2}
\end{eqnarray}
Here we should make two important remarks. First, the trick played in Eq. (\ref{repmeth2}) may not be unique; there could be other recipes to determine the 
von Neumann entropy. What is important is that Hausdorff's moment problem guarantees that these other recipes would give the same result as Eq. (\ref
{repmeth2}). Second, extra care should be taken regarding the following fact. During the calculation of $\textrm{tr}(\hat{\rho}^n)$, we consider $n$ to be a natural number in $\mathbb{N}^*$. We only 
consider $n$ to be real in order to apply the step in Eq. (\ref{repmeth2}), but this does not imply that $\textrm{tr}(\hat{\rho}^x)$, with $x \in \mathbb{R}$, is found by 
simply substituting the natural variable $n$ with the real variable $x$. What Hausdorff's moment problem guarantees is that, in principle, $\textrm{tr}(\hat{\rho}^x)$ 
could be uniquely determined from $\textrm{tr}(\hat{\rho}^n)$ with some proper recipe.  

\subsubsection{Examples from classical probability theory}

To wrap it up, to calculate the von Neumann entropy using the replica method, we have to find $\textrm{tr}(\hat{\rho}^n)$ as a function of 
$n \in \mathbb{N}^*$, and then we need to calculate the derivative of this quantity with respect to $n$ at $n=1$.
For the sake of completeness, let us consider two examples from elementary classical probability theory which illustrate very well the validity of the replica method. These examples have no immediate connection with Hausdorff's moment problem but we find it instructive to see that the replica method works for classical distributions as well. 

First, consider the Gaussian distribution over some real variable $x$ with mean value $\mu$ and standard deviation $\sigma$,
\begin{eqnarray}
P^{(G)}(x)=\frac{1}{\sigma \sqrt{2\pi}} e^{-\frac{(x-\mu)^2}{2 \sigma^2}}.
\label{gaussian}
\end{eqnarray}
The entropy of this distribution can be found by calculating the moments, which are
\begin{eqnarray}
m^{(G)}_n=\frac{1}{(2\pi)^{\frac{n-1}{2}} \sqrt{n} \sigma^{n-1} }
\label{gaussianMom}
\end{eqnarray}
and then by finding the derivative of Eq. (\ref{gaussianMom}) with respect to $n$ at $n=1$,
\begin{eqnarray}
-\frac{dm^{(G)}_n}{dn}\Bigg|_{n=1}=\frac{1}{2} \ln (2 e \pi \sigma^2)
\end{eqnarray}
which is indeed the well-known entropy of the Gaussian distribution.

Now, let us consider the Poisson distribution, 
\begin{eqnarray}
P^{(P)}(k)=\frac{\lambda^k}{k!}e^{-\lambda},\ \lambda>0,\ k\ \in\ \mathbb{N}.
\label{poisson}
\end{eqnarray}
The moments read,
\begin{eqnarray}
m^{(P)}_n=\sum_{k=0}^{\infty} \frac{\lambda^{n k}}{k!^n}e^{-n \lambda}
\end{eqnarray}
and from the latter we find,
\begin{eqnarray}
-\frac{dm^{(P)}_n}{dn}\Bigg|_{n=1}=\lambda(1-\ln\lambda)+e^{-\lambda}\sum_{k=0}^{\infty} \frac{\lambda^k}{k!} \ln k!
\end{eqnarray}
which is indeed the entropy of the Poisson distribution. This result reminds us the simple fact that a non-summation expression for the 
entropy is not always available, even for a simple classical distribution. As we see in this paper, a similar situation prevails when considering the output entropy of a parametric amplifier that is fed with an arbitrary Fock state $|m\rangle$ with $m>1$.

\subsection{2. Entropy produced by amplifying a Fock state}

In the main text, we consider the amplification of an arbitrary Fock state $|m\rangle$. It is shown that the corresponding output state 
can be written as
\begin{eqnarray}
\hat{\rho}_m = \sum_{k=0}^\infty  (1-|\tau|^2)^{m+1} \binom{k+m}{k} |\tau|^{2k}  \,  |k\rangle\langle k|
\end{eqnarray}
so that we obtain a closed expression
\begin{eqnarray}
\textrm{tr}(\hat\rho^n_m )=\frac{(1-|\tau|^2)^{n(m+1)}}{|\tau|^{2n}} \, \textrm{Li}_{-n}^{(m)}(|\tau|^{2n})
\label{trmn2}
\end{eqnarray}
where
\begin{eqnarray}
\textrm{Li}_{-n}^{(m)}(\zeta)  \equiv  \sum_{k=0}^{\infty} \binom{k+m}{k}^n \zeta^{k+1} .
\end{eqnarray}
The function $\textrm{Li}^{(1)}_{-n}(\zeta)$ is known as the polylogarithm of order $-n$. It is connected with the Eulerian polynomials in the following way, 
\begin{eqnarray}
\textrm{Li}^{(1)}_{-n}(\zeta)=\frac{1}{(1-\zeta)^{n+1}}\sum_{k=0}^{n-1} E (n,k) \, \zeta^{k+1}
\label{Li1}
\end{eqnarray}
where
\begin{eqnarray}
E (n,k) \equiv \sum_{j=0}^k \binom{n+1}{j}(-1)^j (k-j+1)^n
\label{eulerian}
\end{eqnarray}
are the Eulerian numbers.

The entropy $S_m$ of the output state $\hat{\rho}_m$ is obtained by taking the derivative of Eq.~(\ref{trmn2}) with respect to $n$, 
taking into account that $\textrm{Li}_{-1}^{(m)}(\zeta)=\zeta/(1-\zeta)^{m+1}$, which yields
\begin{eqnarray}
\lefteqn{   S_m=-\frac{\partial}{\partial n} \textrm{tr}(\hat\rho^n_m)\Big|_{n=1}=\ln \frac{|\tau|^2 ~~~}{(1-|\tau|^2)^{m+1}}     }\nonumber \\
& ~~~~~~~~~~~   -\frac{(1-|\tau|^2)^{m+1}}{|\tau|^2 ~~~}\frac{\partial}{\partial n} \, \textrm{Li}_{-n}^{(m)}(|\tau|^{2n})\Big|_{n=1}.
\label{Sm2}
\end{eqnarray}
This can be written more explicitly as
\begin{eqnarray}
\nonumber S_m&=& (m+1) \, S_0 +\ln m!-\left(1-|\tau|^2\right)^{m+1}\\
&\times & \sum_{k=0}^\infty \binom{k+m}{k} |\tau|^{2k} \ln \frac{(k+m)!}{k!} .
\label{Sm3}
\end{eqnarray}

Closed expressions of $S_m$ are hard to extract for $m>1$ since the function $\textrm{Li}_{-n}^{(m)}(\zeta)$ cannot be written in a non-summation form for $m>1$. This is not so unexpected as it is similar to the case of a Poisson distribution, see Section 1.
Note that $S_m$ can also be calculated analytically using the standard definition of the entropy, Eq.~(\ref{vne}), but the replica method provides an alternative way to achieve this calculation which is straightforward and remains applicable even when Eq.~(\ref{vne}) cannot be exploited, see Section 3.

\subsection{3. Entropy produced by amplifying a superposition of the vacuum and a Fock state}

\subsubsection{Calculation of $\textrm{tr}(\hat{\rho}^n)$}

In the main text, it is shown that if the input state is a superposition $|\psi\rangle = (|0\rangle+z |m\rangle)/\sqrt{1+z^2}$, then the output state $\hat{\rho}$ is such that
\begin{eqnarray}
\textrm{tr}(\hat{\rho}^n)=\frac{\textrm{tr}(\hat\rho_0^n )}{(1+z^2)^n} \, \Pi_{\partial \lambda}(n)\exp\left(\bar{\lambda}^\dagger N \bar{
\lambda}\right)\Big |_{\bar{\lambda}=\bar{0}} ,
\label{trnPi}
\end{eqnarray}
where $\textrm{tr}(\hat\rho_0^n )$ corresponds to a vacuum input state, i.e., $z=0$.
Here, we define the matrix $N=(1-|\tau|^{2n})^{m} M^{-1}$, with
\begin{eqnarray}
 M^{-1} = \frac{1}{1-|\tau|^{2n}}
\begin{pmatrix}
1                         & |\tau|^2                    & \ldots                    & |\tau|^{2(n-1)}\\
|\tau|^{2(n-1)}           & 1                           & \ldots                    & |\tau|^{2(n-2)}\\
\vdots                    & \vdots                      & \ddots                    & \vdots\\
|\tau|^2                  & |\tau|^4                    & \ldots                    & 1
\end{pmatrix}
\label{C}
\end{eqnarray}
being a circular matrix. The differential operator $\Pi_{\partial \lambda}(n)$ has the form
\begin{eqnarray}
\Pi_{\partial \lambda}(n)=\sum_{k=0}^{n} c^{2k} \, \Pi_{2k}(n)
\label{Pi}
\end{eqnarray}
where each $\Pi_{2k}(n)$ contains $\binom{n}{k}^2$ terms that give a non-zero result when 
acting on $\exp\left(\bar{\lambda}^\dagger N \bar{\lambda}\right)$ and taking the value at $\bar{\lambda}=\bar{0}$. 
These terms are all the derivatives of even order $(2,\ 4,\ \ldots, 2n)$ with respect to $\lambda$ such that the number of $\lambda$ is equal to the 
number of $\lambda^*$ for each derivative. For example, by keeping terms that return non-zero result in Eq. (\ref{trnPi}), we have
\begin{eqnarray}
\nonumber  k=0&:&\ \Pi_{0}(n)=1\\
\nonumber  k=1&:&\ \Pi_{2}(n)=\frac{\partial^{2 m}}{\partial \lambda_1 \partial \lambda_1^*}+\ldots\\
\nonumber  k=2&:&\ \Pi_{4}(n)=\frac{\partial^{4 m}}{\partial \lambda_1 \partial \lambda_1^* \partial \lambda_2 \partial \lambda_2^*}+\ldots\\
\nonumber  &\vdots&\\
k=n&:&\ \Pi_{2n}(n)=\frac{\partial^{2 m n}}{\partial \lambda_1 \partial \lambda_1^*\ldots \partial \lambda_n \partial \lambda_n^*}
\label{eachPi}
\end{eqnarray}

It can be verified that the term with $k=n$ gives, when acting on the exponential of Eq. (\ref{trnPi}),
\begin{eqnarray}
\Pi_{2n} \exp\left(\bar{\lambda}^\dagger N \bar{\lambda}\right)\Big |_{\bar{\lambda}=\bar{0}}=
m!^n \frac{1-|\tau|^{2n}}{|\tau|^{2n}} \, \textrm{Li}_{-n}^{(m)}(|\tau|^{2n})
\label{last}
\end{eqnarray}
where $\textrm{Li}_{-n}^{(m)}(|\tau|^{2n})$ is defined in the main text. In other words, this term is connected with the entropy $S_m$ when the input state is the Fock state $|m\rangle$, something that can be seen by taking the limit $z\rightarrow \infty$ in Eq. (\ref{trnPi}). Also, it is not difficult to see that for each operator with $k=0,1\ldots,n$ in 
Eq. (\ref{eachPi}), there are exactly $\binom{n}{k}$ terms where the derivatives with respect to conjugate pairs of $\lambda$'s appear. For example, such a term is $\partial^{6}/\partial \lambda_1\partial \lambda_1^*\partial \lambda_2\partial \lambda_2^*\partial \lambda_3\partial \lambda_3^*$. These terms will give 
a result with no dependence on $|\tau|$ in the numerator when it acts on the exponential of Eq. (\ref{trnPi}). If we extract all these terms and gather them 
together, substituting $c$ with its definition 
\begin{eqnarray}
c=\frac{z}{\sqrt{m!}}(1-|\tau|^2)^{m/2},
\end{eqnarray}
we can write
\begin{eqnarray}
\nonumber   \sum_{k=0}^n \binom{n}{k} z^{2k} \Bigg(\frac{1-|\tau|^2}{1-|\tau|^{2n}}\Bigg)^{m k}=\Bigg[1+  z^{2} \Bigg(\frac{1-|\tau|^2}{1-|\tau|^{2n}}\Bigg)^m \Bigg]^n.  \\
\label{constTerms}
\end{eqnarray} 
From all this, we obtain the expression
\begin{eqnarray}
\nonumber \lefteqn{    \textrm{tr}(\hat{\rho}^n)=\frac{\textrm{tr}\hat\rho_0^n}{(1+z^2)^n}\Bigg\{ \Bigg[1+  z^{2} \Bigg(\frac{1-|\tau|^2}{1-|\tau|^{2n}}\Bigg)^m \Bigg]^n    ~~~ }  \\
&& -z^{2n}\Bigg(\frac{1-|\tau|^2}{1-|\tau|^{2n}}\Bigg)^{m n}+F^{(m)}(n)  +z^{2n}\frac{\textrm{tr}\hat\rho_m^n}{\textrm{tr}\hat\rho_0^n}\Bigg\}  
\label{trnFinal}
\end{eqnarray}
where $\hat\rho_m$ is the output state resulting from an input state $|m\rangle$. In Eq. (\ref{trnFinal}), we subtracted the term proportional to $z^{2n}$ 
as we have used it twice; one in the first term of the form $[\ldots]^n$ and a second time for the very last term. In the same 
equation, $F^{(m)}(n)$ gathers all terms except for the first and last one of Eq. (\ref{trnPi}), but without returning any term with no dependence on $|\tau|$ in 
the numerator, something that we denote as $\tilde\Pi_{2k}(n)$ in the expression
\begin{eqnarray}
\nonumber \lefteqn{   F^{(m)}(n)=\sum_{k=1}^{n-1} \frac{z^{2k}}{m!^k} \Bigg(\frac{1-|\tau|^2}{1-|\tau|^{2n}}\Bigg)^{m k}    } \\ 
&& ~~~ \times \tilde\Pi_{2k}(n) \exp\left(\bar\lambda^\dagger N \bar\lambda \right). 
\label{F}
\end{eqnarray}

\subsubsection{Calculation of $F^{(m)}(n)$ and its derivative}

In the main text, it is shown that by taking the derivative of Eq.~(\ref{trnFinal}) with respect to $n$ and keeping the value at $n=1$, we get
\begin{eqnarray}
S(z)=\frac{1}{1+z^2}S_0+\frac{z^2}{1+z^2}S_m-\frac{\partial}{\partial n}F^{(m)}(n)\Big|_{n=1}.
\label{SwithF}
\end{eqnarray}
We shall now prove that the last term in the right hand side of Eq. (\ref{SwithF}) is equal to zero.
It can be found that Eq. (\ref{F}) assumes the form,
\begin{eqnarray}
\nonumber F^{(m)}(n)&=&\sum_{k=0}^{n-1} \frac{z^{2k}}{m!^k} \Bigg(\frac{1-|\tau|^2}{1-|\tau|^{2n}}\Bigg)^{m k}\times\\
 && \times \sum_{l=0}^{(n-1)k} A_k(n,l)|\tau|^{2(l+m-1)}
\label{F2}
\end{eqnarray}
where $A_k(n,l)$ are unknown coefficients satisfying the constraints,
\begin{eqnarray}
\nonumber A_k(1,l)&=&0\\
A_0 (n,l)&=&0.
\label{cons}
\end{eqnarray}
Now, Eq. (\ref{F2}) may be written,
\begin{eqnarray}
F^{(m)}(n)=\sum_{k=0}^{n-1} R^{(m)}(n,k).
\label{F3}
\end{eqnarray}
Using the Euler-McLaurin summation formula \cite{apostol} we get,
\begin{eqnarray}
\nonumber F^{(m)}(n)&=&\mathop{\int} \limits_{0}^{n-1} dx R^{(m)}(n,x)+\\
\nonumber &&+\frac{1}{2}\Big[R^{(m)}(n,n-1)+R^{(m)}(n,0)\Big]+\\
\nonumber &&+\sum_{r=1}^p \frac{B_{2r}}{(2r)!}\Big[R^{(m)(2r-1)}(n,n-1)-\\
&&-R^{(m)(2r-1)}(n,0)\Big]+\textrm{Rem}
\label{EulerSum}
\end{eqnarray}
where $B_{2r}$ are the Bernoulli numbers and we symbolize as $R^{(m)(\mu)}(n,a)$ the differentiation with respect to the second argument,
\begin{eqnarray}
R^{(m)(\mu)}(n,a)=\frac{\partial^{\mu}}{\partial x^{\mu}}R^{(m)}(n,x)\Bigg|_{x=a}.
\end{eqnarray}
Differentiation with respect to the first argument will be denoted explicitly as $\partial_n R^{(m)}(n,x)$. The last term in Eq.~(\ref{EulerSum}) is the remainder 
and has the form,
\begin{eqnarray}
\textrm{Rem}=-\mathop{\int}  \limits_{0}^{n-1} dx \frac{P_{2p}(x)}{(2p)!}R^{(m)(2p)}(n,x).
\label{rem}
\end{eqnarray}
For $x>0$ we get the periodic Bernoulli functions $P_n(x)=B_n(x-[x])$, where $[x]$ is the largest integer $x$, while $P_n(0)=B_n$.

If we perform the derivative of Eq. (\ref{EulerSum}) with respect to $n$ at $n=1$, we see, taking Eq. (\ref{cons}) into account, that all terms vanish. This is 
because three kind of terms appear,
\begin{eqnarray}
R^{(m)}(1,0)=0\\
\mathop{\int} \limits_0^0 dx (\dots)=0\\
\partial_n R^{(m)}(n,0)\Bigg|_{n=1}=\partial_n 0=0\\
R^{(m)(2r)}(1,0)=0.
\end{eqnarray} 
Thus, we conclude that
\begin{eqnarray}
\frac{d}{dn}F^{(m)}(n)\Big|_{n=1}=0.
\label{dF0}
\end{eqnarray}
as advertised.

\end{document}